\documentclass[fleqn,10pt]{wlscirep}
\usepackage[utf8]{inputenc}
\usepackage[T1]{fontenc}
\newlength{\sizeonethird}
\usepackage{setspace}
\begin{document}
\title{Electronic mechanism of sub-100-fs demagnetization induced by a femtosecond light pulse}
\author[1,2,*]{Konrad J. Kapcia}
\author[3,2]{Victor Tkachenko}
\author[4]{Flavio Capotondi}
\author[3,5]{Alexander Lichtenstein}
\author[3,6,7]{Serguei Molodtsov}
\author[8]{Przemys\l{}aw Piekarz}
\author[2,8]{Beata Ziaja}
\affil[1]{Institute of Spintronics and Quantum Information, Faculty of Physics and Astronomy, Adam Mickiewicz University in Pozna\'n, Uniwersytetu Pozna\'{n}skiego 2, 61614 Pozna\'{n}, Poland}
\affil[2]{Center for Free-Electron Laser Science CFEL, Deutsches Elektronen-Synchrotron DESY, Notkestr. 85, 22607 Hamburg, Germany}
\affil[3]{European XFEL GmbH, Holzkoppel 4, 22869 Schenefeld, Germany}
\affil[4]{Elettra-Sincrotrone Trieste S.C.p.A, 34149 Trieste, Basovizza, Italy}
\affil[5]{University of Hamburg, Jungiusstr.  9, 20355 Hamburg, Germany}
\affil[6]{Institute of Experimental Physics, TU Bergakademie Freiberg, Leipziger Strasse 23, 09599 Freiberg, Germany}
\affil[7]{Center for Efficient High Temperature Processes and Materials Conversion (ZeHS), TU Bergakademie Freiberg, Winklerstrasse 5, 09599 Freiberg, Germany}
\affil[8]{Institute of Nuclear Physics, Polish Academy of Sciences,  Radzikowskiego 152, 31-342  Krak\'ow, Poland}
\affil[*]{Correspondence should be addressed to konrad.kapcia@amu.edu.pl (K.J.K.).}
%
%
%
\begin{abstract}
A quantitative understanding of the processes that trigger light-induced demagnetization on ultrashort timescales is crucial for achieving an ultrafast, radiation-controlled magnetic response in materials. This milestone is essential for developing next-generation magnetic storage devices and ultrafast magnetic switches. In this theoretical study, we investigated demagnetization triggered in a single magnetic domain by light pulses ranging from a few to a few tens of femtoseconds in duration, with photon energies spanning the optical and X-ray regimes, under strongly non-equilibrium conditions. We predicted a loss of magnetization in the sub-100-fs range in all cases, primarily due to the excitation of the electronic system and the subsequent redistribution of electrons within the magneto-sensitive band. The considered timescales were too short for phonon-mediated processes or inter-site Heisenberg exchange processes to contribute significantly. These findings pave the way for highly accurate, radiation-driven magnetization control in magnetic materials at sub-100-femtosecond timescales with potential practical applications.
\end{abstract}
\flushbottom
\maketitle
\section*{Introduction}

Ultrafast radiation-induced demagnetization of solids have been studied since its discovery in 1996 \cite{BeaurepairePRL1996}, mostly with lasers working in the optical wavelength regime \cite{KirilyukRMP2010,KoopmansNatMat2010,PfauNatCom12,SanderJPhysD17}. Since a decade, ultrafast
demagnetization can also be triggered with X-ray radiation emitted by free-electron lasers (FELs) that  generate intense, coherent pulses of femtosecond duration and tunable wavelength \cite{AckermannNatPhot2007,EmmaNatPhot2010,PileNatPhot2011,AllariaNatPhot2012}. 
Interestingly, the timescales of observed demagnetization both in case of femtosecond XUV or optical lasers are also in the femtosecond regime \cite{GuttPRB10,WangPRL2012,PfauNatCom12,MullerPRL2013a,WuPRL2016,WillemsStrDyn2017,ChenPRL2018,SchneiderPRL2020,StammNatMat07,HennesAppLSci2021}.

This radiation-induced and radiation-controlled magnetic response of irradiated solid materials can be a milestone for constructing next-generation magnetic storage devices and ultrafast magnetic switches. Prototype studies on optical magnetic switches on nanometer length scales  have been already performed \cite{YaoNanoLett2022,CiuciulkaitePRMaterials2020}. For the practical implementation, it is important that such processes can be steered by a table top photon source. Compact optical lasers and table-top high-harmonic generation sources in XUV regime offer such possibilities.

Many experiments have been performed so far to investigate radiation-induced demagnetization, using various methods for the detection of transient magnetic states such as X-ray magnetic circular dichroism (XMCD) spectroscopy \cite{CarvaEPL2009,StammNatMat07,RosnerStructDyn2020}, small-angle magnetic X-ray scattering (mSAXS) \cite{MullerPRL2013,GuttPRB2009}, photoemission spectroscopy \cite{VaterlausPRL1991,EichSciAdv2017} and magneto-optical Kerr effect (MOKE) \cite{BeaurepairePRL1996,KoopmansNatMat2010,ProbstPRResearch2024}.  
Many theoretical models have been proposed \cite{BeaurepairePRL1996,KoopmansNatMat2010,SchellekensPRL2013,StohrPRL2015,CarpenePRB2015,KorniienkoPRR2024,Murphy2014,Beyerlein2018,Ho2020,Hourahine2020}  as well, in particular, those inspired by the Elliot-Yaffet model \cite{SteiaufPRB2009,CarvaPRL2011,BaralNJP2016, PRB-80-180407-2009} or the time-dependent density functional theory \cite{KriegerJCTC2015,TurgutPRB2016,ChenSC2019,AcharyaPRL2020,BarrosEPJB2020,ScheidNanoLett2021,HarisLeeSciAdv2024,MrudulPRB2024}. 
Some of them also proposed to couple the magnetic dynamics with spin quasi-ballistic transport \cite{BattiatoPRL2010}. However, it has been so far not clear which processes trigger ultrafast magnetization changes in  irradiated magnetic materials after a stimulus by an ultrashort pulse ranging from a few to a few tens of femtoseconds in duration.

In a series of recent papers \cite{KapciaNPJ2022,KapciaPRB2023,KapciaSciRep2024,KapciaCCC2024,AntunesCCP2026}, we have investigated predominant mechanisms for single-domain demagnetization following the fast  XUV and soft X-ray irradiation. Pulse ranging from a few to a few tens of femtoseconds in duration were considered. For this purpose, we have constructed a dedicated code XSPIN, applying single-domain modeling framework, similar to that in \cite{StonerPTRSLA47,TannousEJP08}. Our model enables the study of demagnetization under strongly non-equilibrium conditions following the impact of such ultrashort pulses.
The XSPIN model simulations could identify electronic excitation as a predominant mechanism of the predicted sub-100-fs demagnetization in a single domain. In particular, it was shown that irradiation with femtosecond X-ray pulses, which rapidly releases many electrons, leads to a rapid rearrangement of the electronic occupation in the magnetically sensitive 3d band. This changes the magnetization of the sample on sub-100-fs timescale. Note that the considered timescales are too short for  phonon-mediated processes or processes related to inter-site Heisenberg exchange to contribute significantly. Therefore, effects associated with the rotation of magnetic moments, such as magnons, chiral magnetic structures, and domain walls, do not need to be considered in the model.

In this paper, we demonstrate the universality of the sub-100-fs demagnetization mechanism using the XSPIN simulation tool. This mechanism generally applies to a magnetic domain after its irradiation with ultrashort photon pulses of energies greater than 2 eV. This finding qualitatively explains the similarity of demagnetization timescales observed after the irradiation of magnetic materials with optical or XUV pulses ranging from a few to a few tens of femtoseconds. This work also provides a new perspective on inducing and controlling ultrafast demagnetization of materials with photons ranging in energy from the optical to the soft X-ray spectrum, with potential practical applications.

\section*{Model and simulation tool}

When photons of XUV or soft X-ray energies are absorbed in a bulk material, they release energetic electrons. These electrons then relax on femtosecond timescales through impact (collisional) ionization until their energy becomes too low to excite further electrons \cite{MedvedevNJP2013,MedvedevPRB2013,MedvedevPRB2017,Medvedevf4open2018,MedvedevSciRep2018,LippProcSPIE17}. Our model accounts for all the processes of electron excitation from deep shells, collisional ionization, and the Auger effect.  The low-energy electrons then constitute the so-called low-energy fraction of excited electrons that occupy the lower part of the conduction band until energy loss processes through electron-lattice energy exchange (such as, e.g., electron-phonon coupling) lead to a full relaxation of the electronic system.  Already in Ref. \cite{TkachenkoPRB16}, we have noticed that independently of the initial excitation mechanism, i.e., the incoming photon energy, the final state of the low-energy excited electrons depends only on the average radiation dose absorbed per atom. This observation will be used here to analyze demagnetization of irradiated samples with our XSPIN model. From now on, we will restrict only to the radiation doses below the structural damage threshold, i.e., we will assume that the radiation dose is so low that it causes only electronic excitation and not nuclei relocations within the irradiated material.

\begin{figure}[t!]
    \centering
    \includegraphics[width=\sizeonethird]{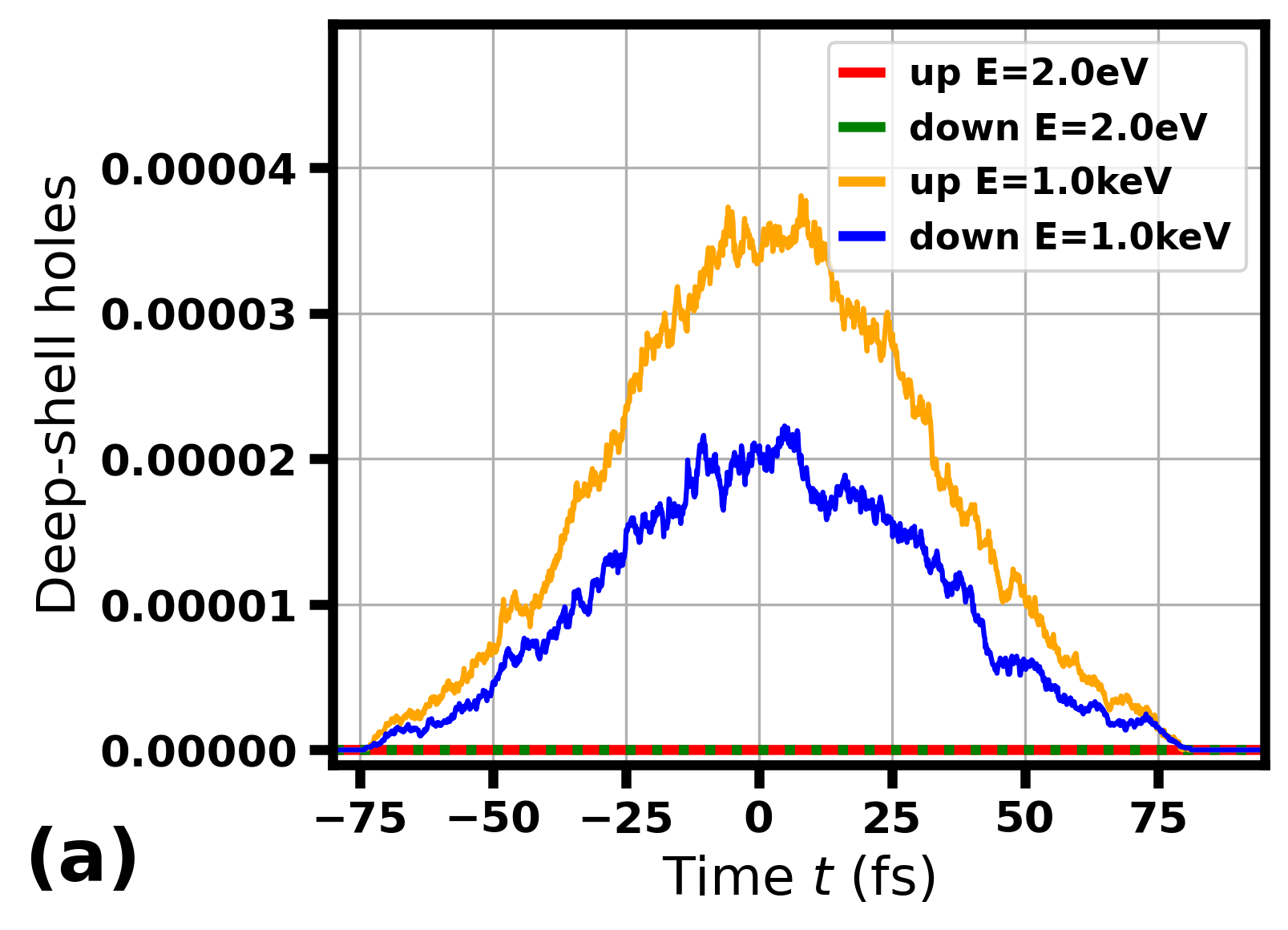}
    \includegraphics[width=\sizeonethird]{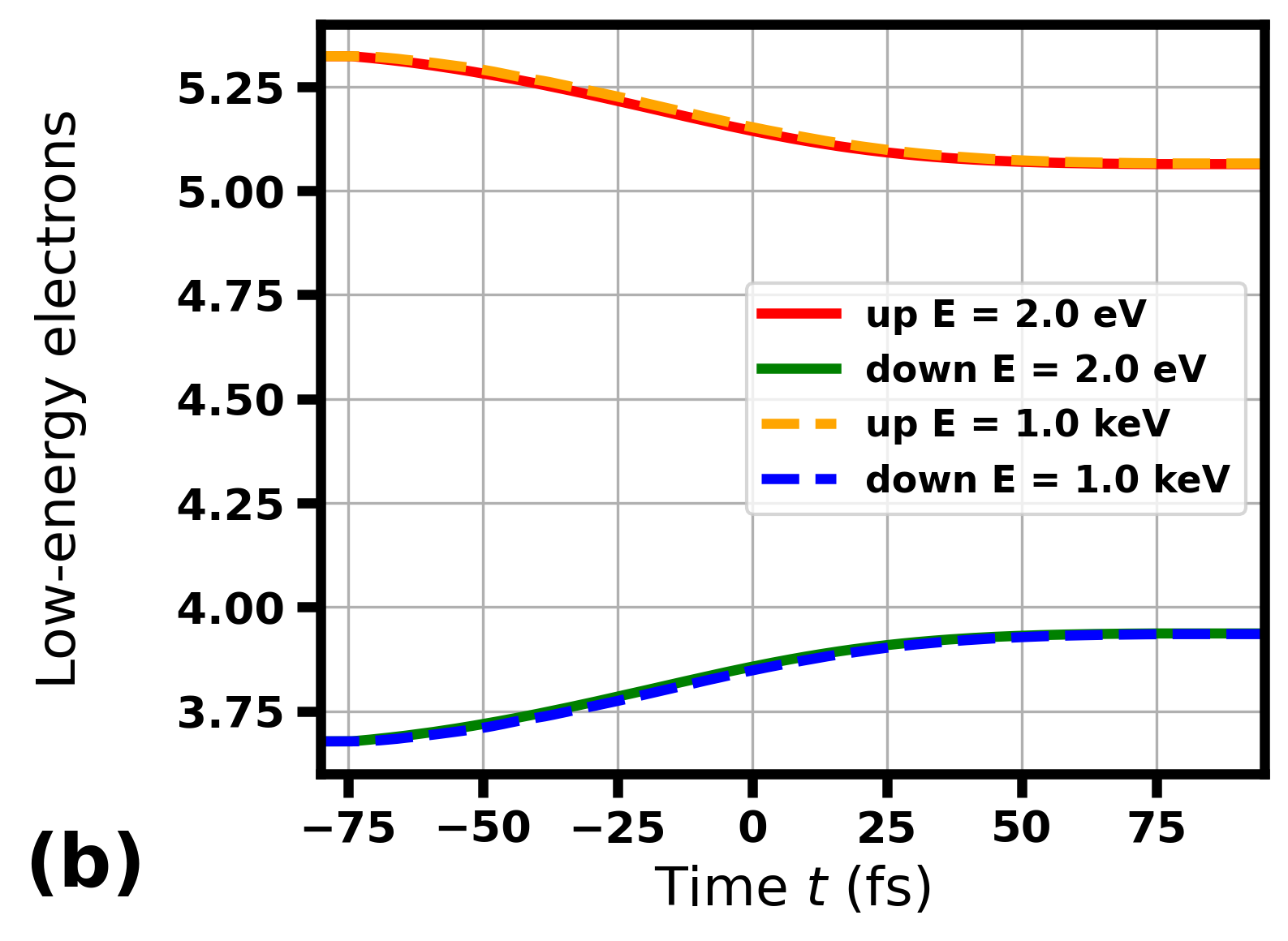}\\
    \includegraphics[width=\sizeonethird]{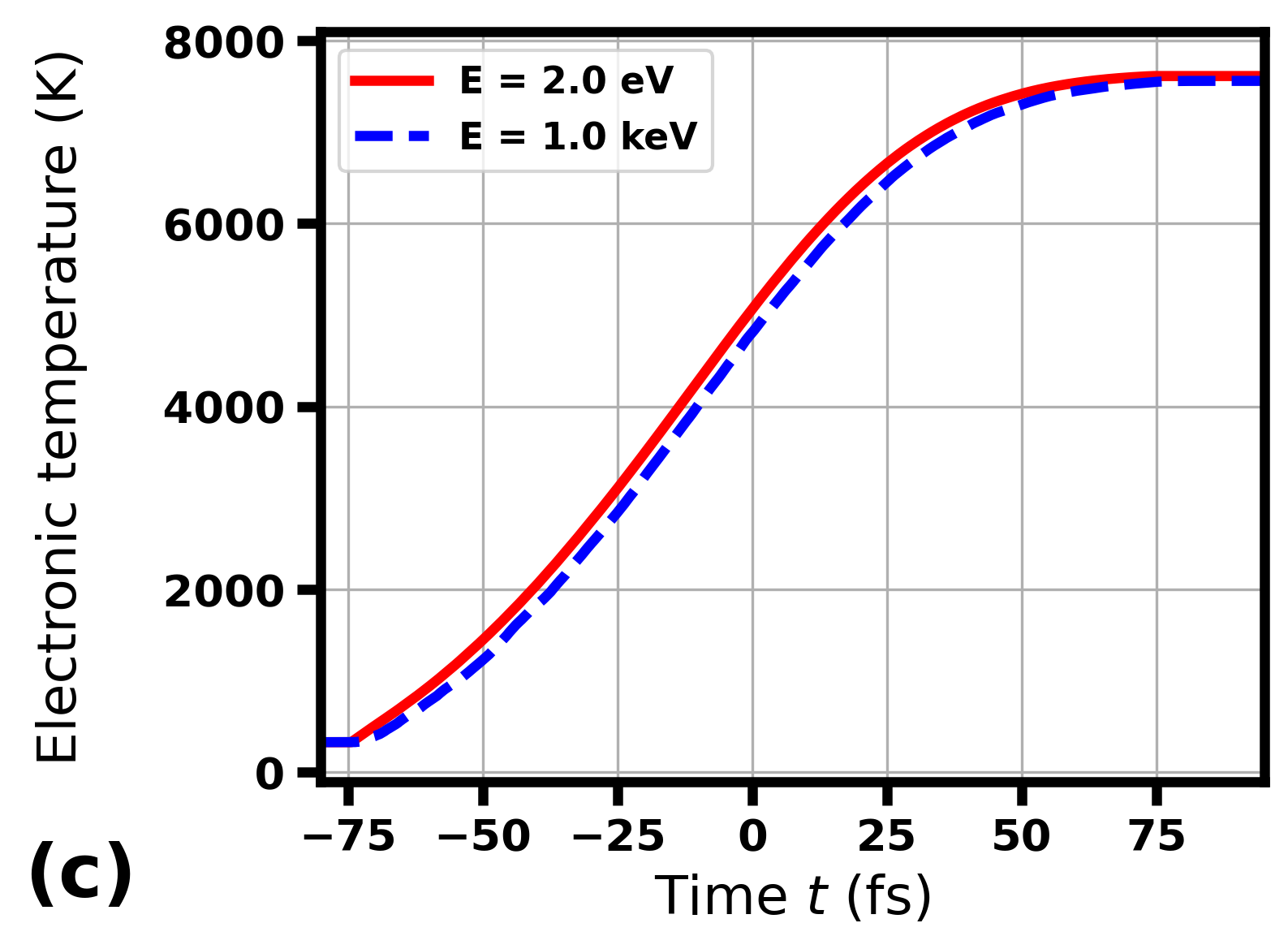}
    \includegraphics[width=\sizeonethird]{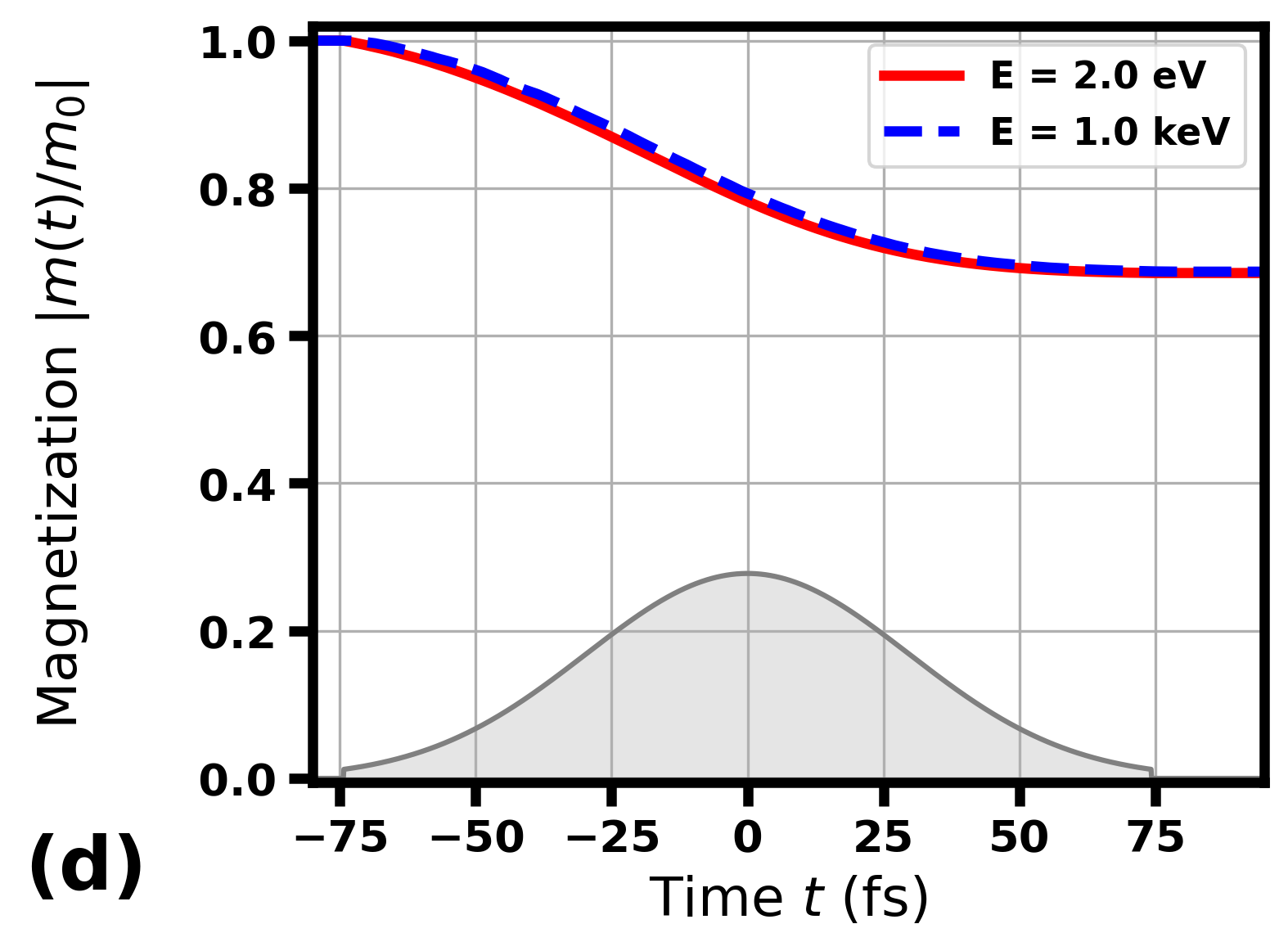}
\caption{\textbf{Transient electronic and magnetic properties of Co predicted by XSPIN in an single magnetic domain under optical irradiation with $2$ eV pulse or X-ray irradiation with $1000$ eV pulse (as labeled), both of $70$ fs FWHM duration.} (a) number of deep-shell holes produced (per atom; spin-resolved), (b) number of low-energy electrons per atom (with energy below the cutoff of $15$ eV, spin-resolved), (c) electronic temperature, (d) magnetization (normalized to its initial value $m_0$ before the pulse), all plotted as a function of time. Temporal profile of the pulse triggering the dynamics is schematically depicted.
}
    \label{fig1}
\end{figure}

\begin{figure}[t!]
    \centering
    \includegraphics[width=\sizeonethird]{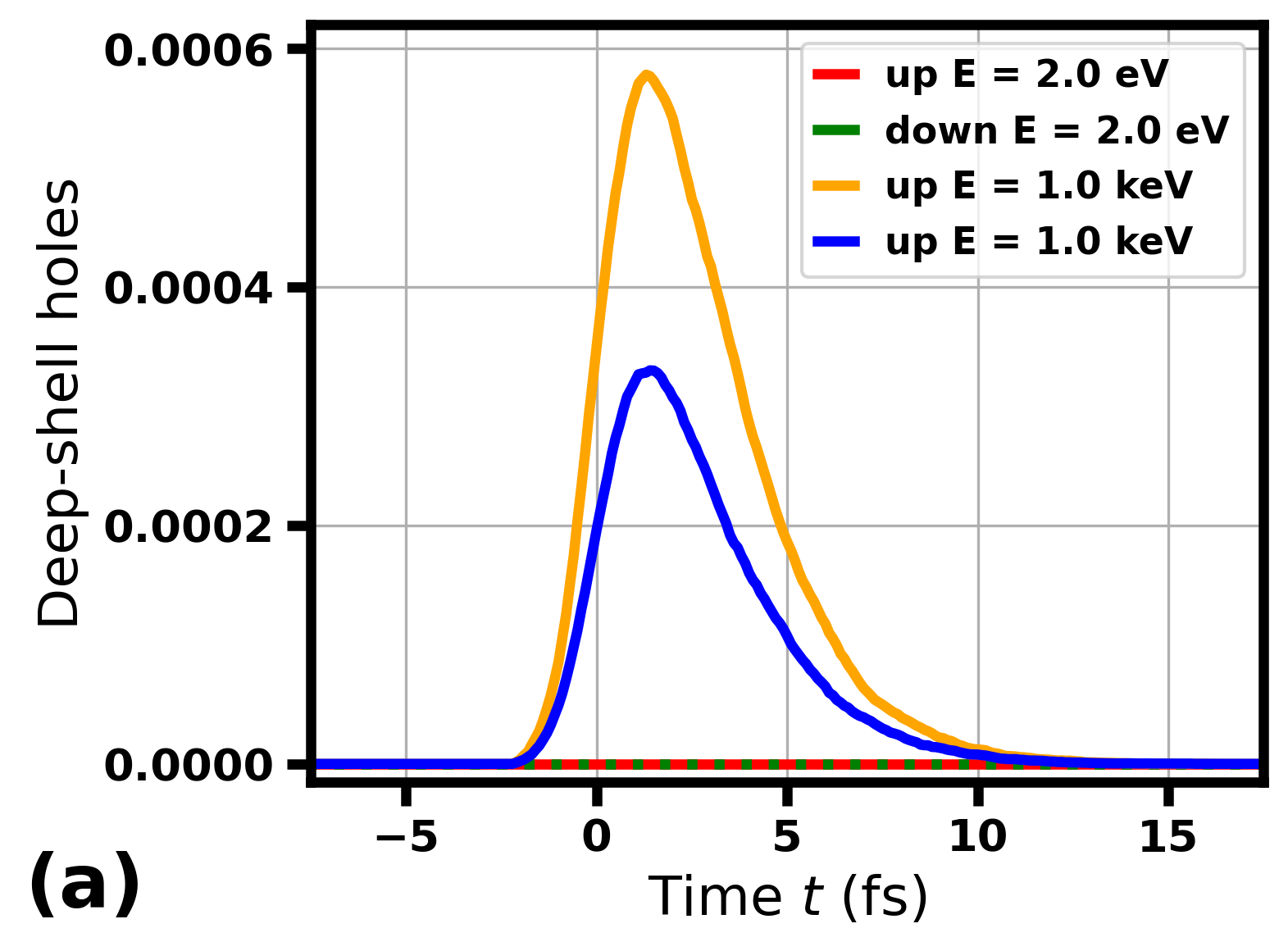}
    \includegraphics[width=\sizeonethird]{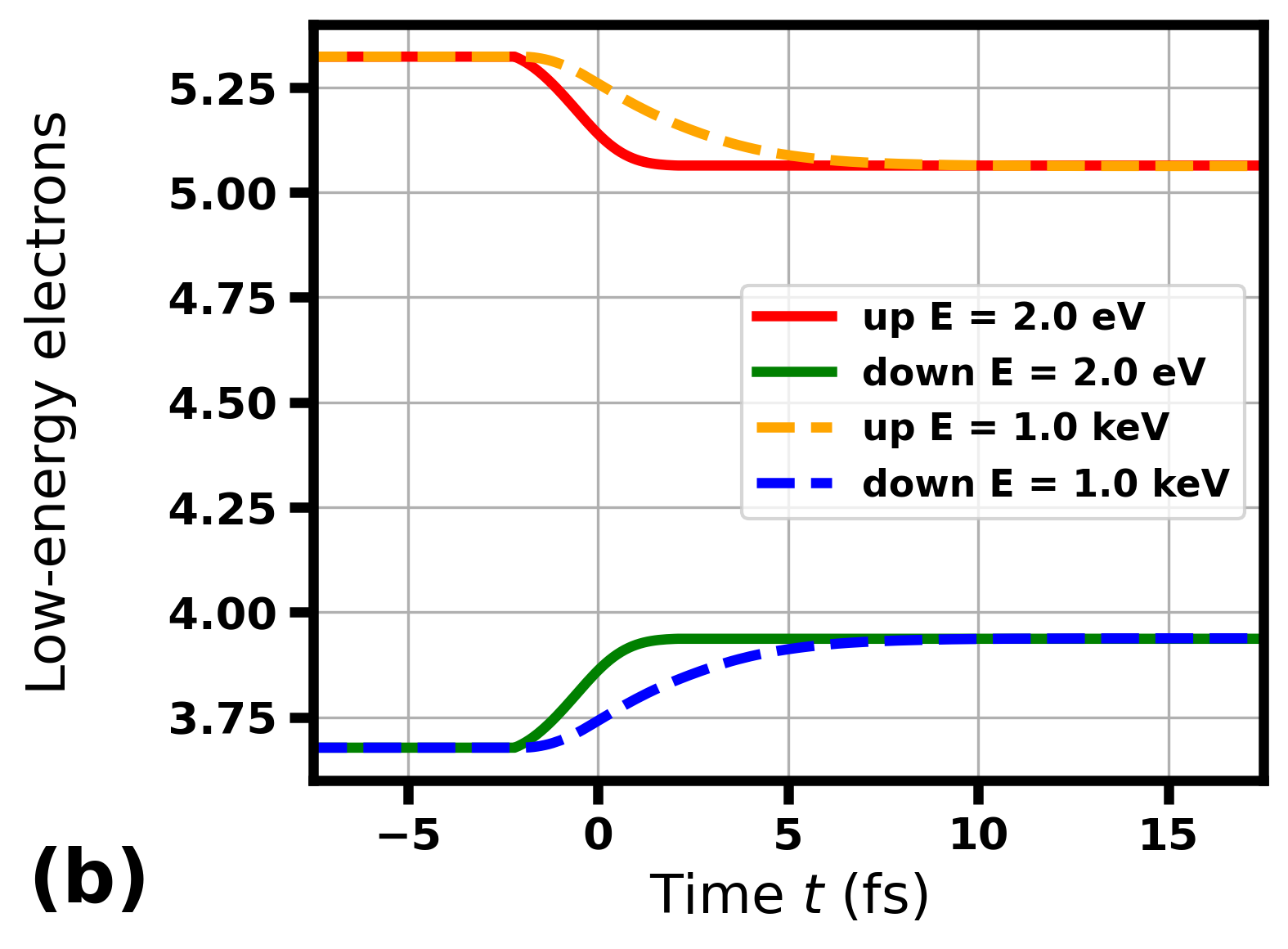}\\
    \includegraphics[width=\sizeonethird]{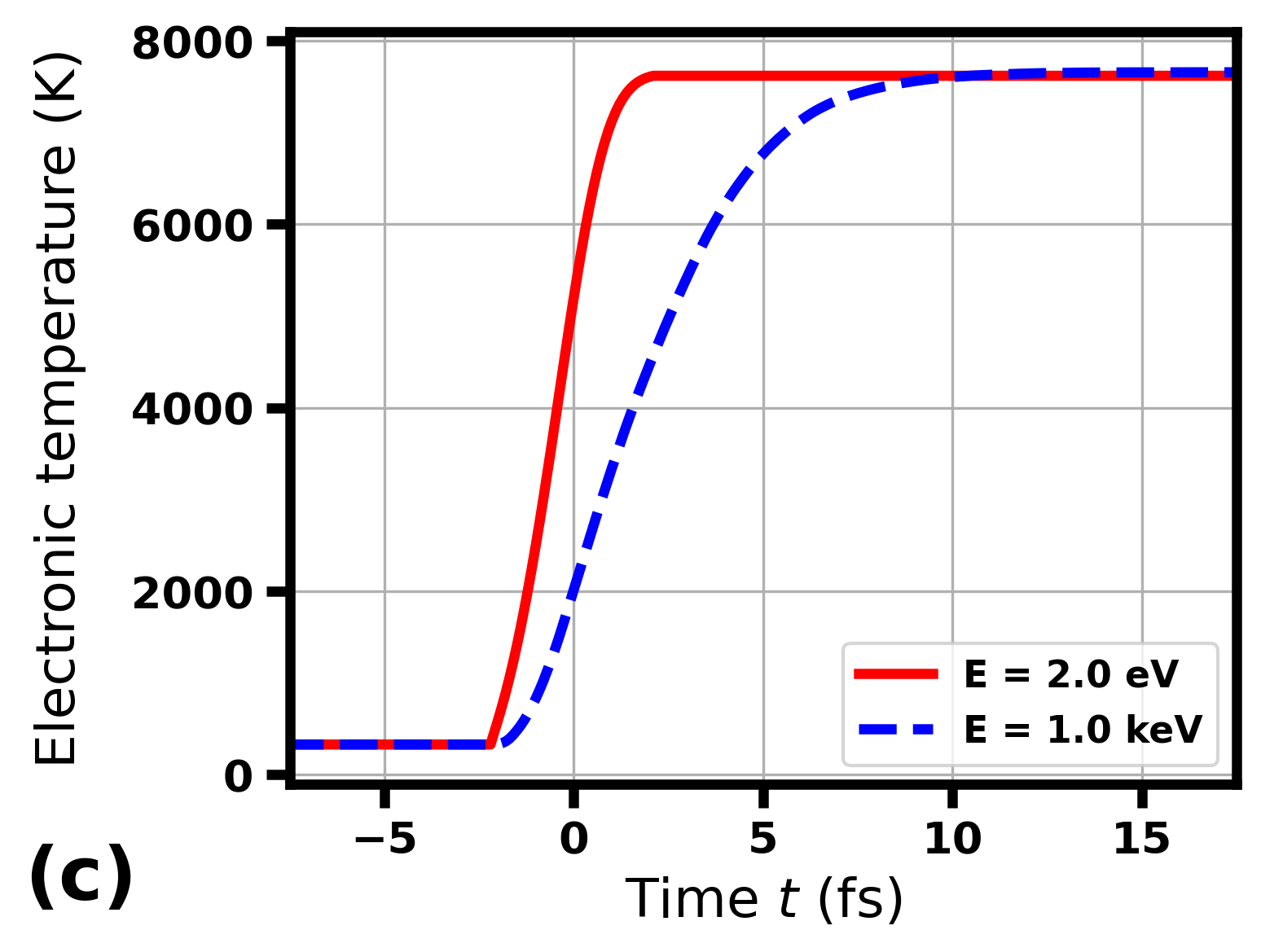}
    \includegraphics[width=\sizeonethird]{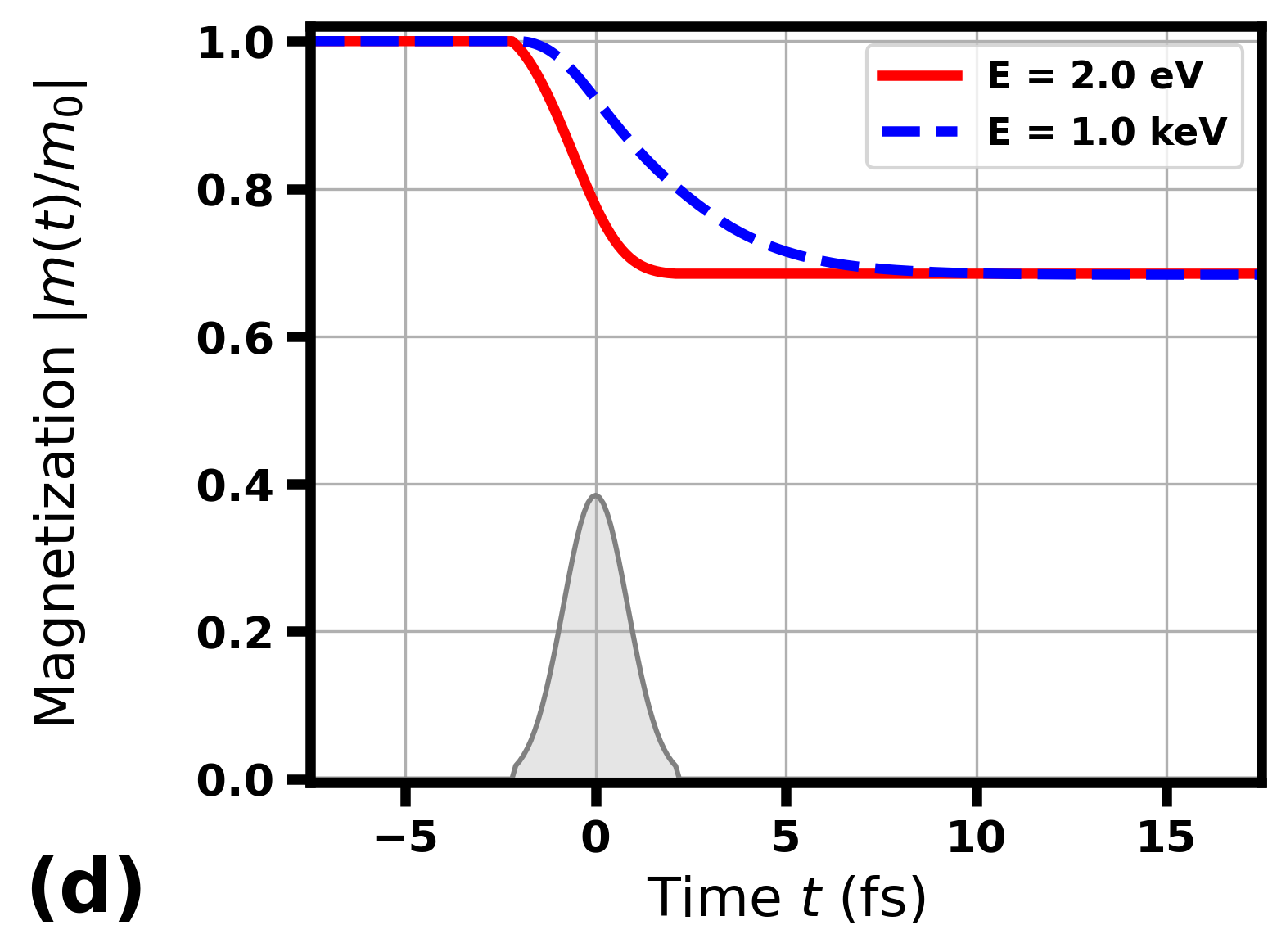}
    \caption{\textbf{Transient electronic and magnetic properties of Co predicted by XSPIN in a single magnetic domain under optical irradiation with $2$ eV pulse or X-ray irradiation with $1000$ eV pulse (as labeled), both of $2$ fs FWHM duration.} Other simulation parameters are as in Fig.~\ref{fig1}.}
    \label{fig2}
\end{figure}

We will perform our study on Co and Ni samples. As was done in studies of irradiated unpolarized materials with the XTANT code (see, for example, \cite{MedvedevPRB2013, Medvedevf4open2018, Lipp2022dftbp, InouePRL2022}), our XSPIN model (see \cite{KapciaNPJ2022, KapciaPRB2023, KapciaSciRep2024, KapciaCCC2024,AntunesCCP2026}) for magnetic samples applies a non-equilibrium electron energy distribution (see \cite{PRL-107-165003-2011, PRL-104-125002-2010}). This distribution consists of a high-energy electron fraction containing high-energy conduction-band electrons treated with the Monte Carlo method and a low-energy fraction containing valence and low-energy conduction-band electrons. The latter is assumed to obey a transient Fermi-Dirac distribution with a transient temperature and chemical potential. At each time step, the energy deposited into the electronic system by incoming photons through various photoabsorption channels is added to the total energy. This increases the transient electronic temperature and changes the transient chemical potential. Following the reasoning presented in \cite{KapciaNPJ2022}, we enforce a common transient electronic temperature and chemical potential for the entire electronic system (i.e., for electron fractions with both spin up and spin down) at each time step. Any change in electronic temperature then induces migration of electrons between spin-up and spin-down fractions, accounting for spin-flip processes (i.e., Hund exchange). This leads to a change in the total polarization of the electronic system. For more details, see the Supplementary Methods.

Consistent with the assumptions used in the modeling tools XTANT (for non-magnetic materials) and XSPIN (for magnetic materials), the band structure in the models only changes when nuclei begin to move. This is not the case here due to the low intensity of the considered irradiation. Electronic excitation and relaxation also do not affect the band structure. The band structure is calculated using the Vienna Ab-initio Simulation Package (VASP) \cite{VASP,VASP1,VASP2,VASP3}. In VASP, projector augmented-wave (PAW) potentials and the generalized gradient approximation (GGA) \cite{BlochPRB94} in the Pardew, Burke, and Ernzerhof (PBE) parameterization  \cite{PerdewPRL96} are used to calculate the electronic density of states for bulk fcc Ni and fcc Co.
The density of states (DOS) obtained for both materials for low energy electrons (Supplementary Fig. S1) correspond to standard density functional theory results, e.g., \cite{Coey2001,GrechnevPRB2007,HafnerPRB2008,LizarragaSciRep2017,HeSciRep2019,ZahnPRR2021} (cf. also \cite{KapciaNPJ2022,KapciaPRB2023,KapciaSciRep2024}). 
Additionally, the visualizations of the DOS with the occupations after the end of the pulse (after $75$~fs) for both materials are presented in Supplementary Results (Supplementary Fig. S2).

We  emphasize that we are aware of many existing models that exploit various mechanisms for radiation-induced demagnetization. For example, see references \cite{SchellekensPRL2013, Nature2020, PhysRevB-108-L060404-2023, PhysRevB-90-144420-2014} and the many others cited in this paper.  However, these mechanisms do not apply under the strongly non-equilibrium conditions of the electronic system considered here on a sub-100-fs timescale.  In particular, the effects of electron-phonon coupling can be neglected because they occur over a much longer timescale (i.e., sub-picoseconds, as discussed in \cite{MedvedevPRB2017, ZiajaSciRep2016}). Additionally, a rough estimation confirms that single-electron processes associated with on-site Hund exchange ($J_H$ is of the order of 1 eV) should dominate. In contrast, processes related to inter-site Heisenberg exchange ($J_{ex}$ is of the order of 0.1 eV) should be of minor importance, as they are about ten times slower ($t\propto\hbar/J$). Therefore, effects associated with the rotation of magnetic moments, such as magnons, chiral magnetic structures, and domain walls, do not need to be considered in our model.  Additionally, XSPIN does not trace the overall angular momentum of the system \cite{KapciaNPJ2022}. 

\begin{figure}[t!]
    \centering
    \includegraphics[width=\sizeonethird]{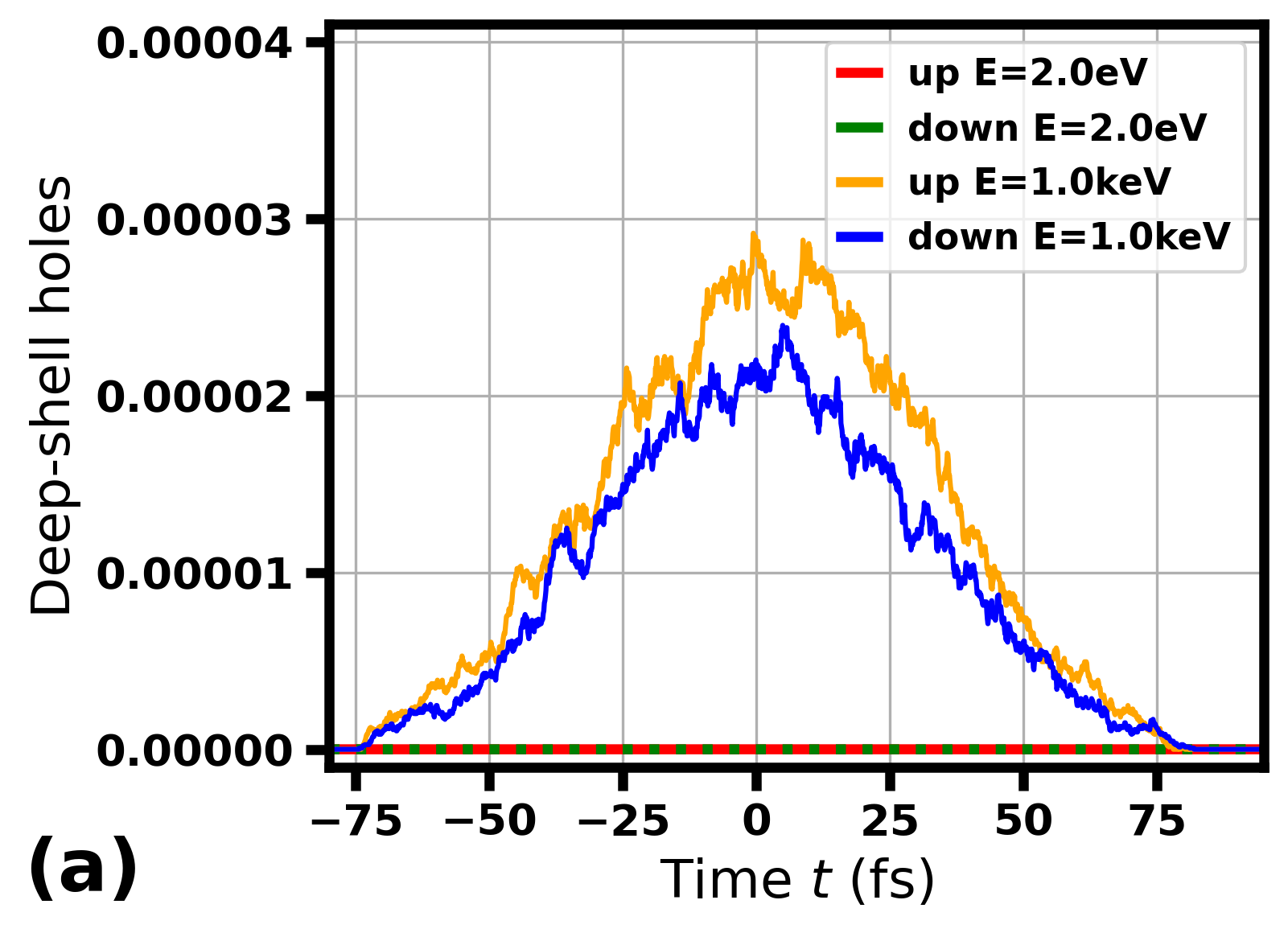}
    \includegraphics[width=\sizeonethird]{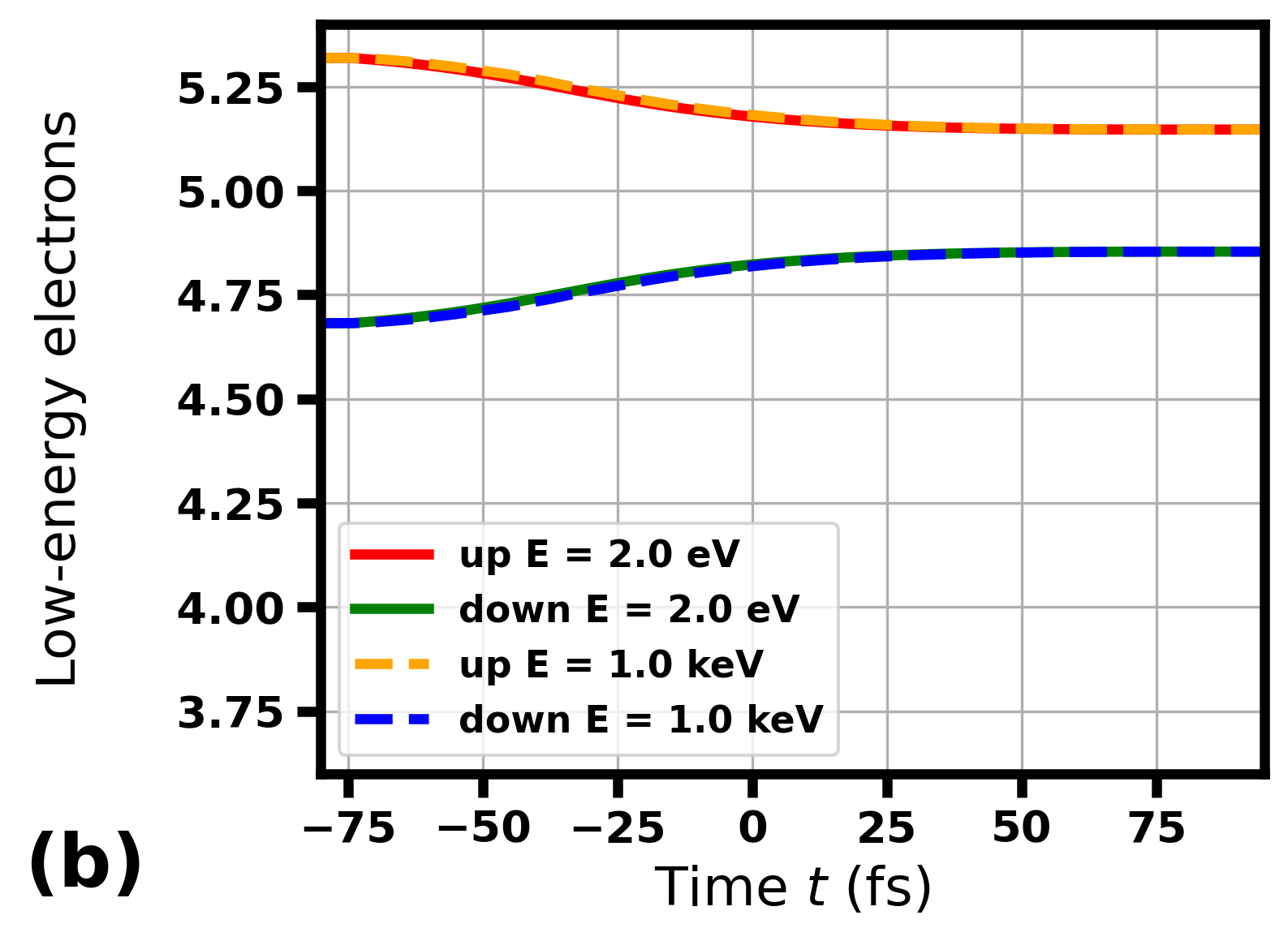}\\
    \includegraphics[width=\sizeonethird]{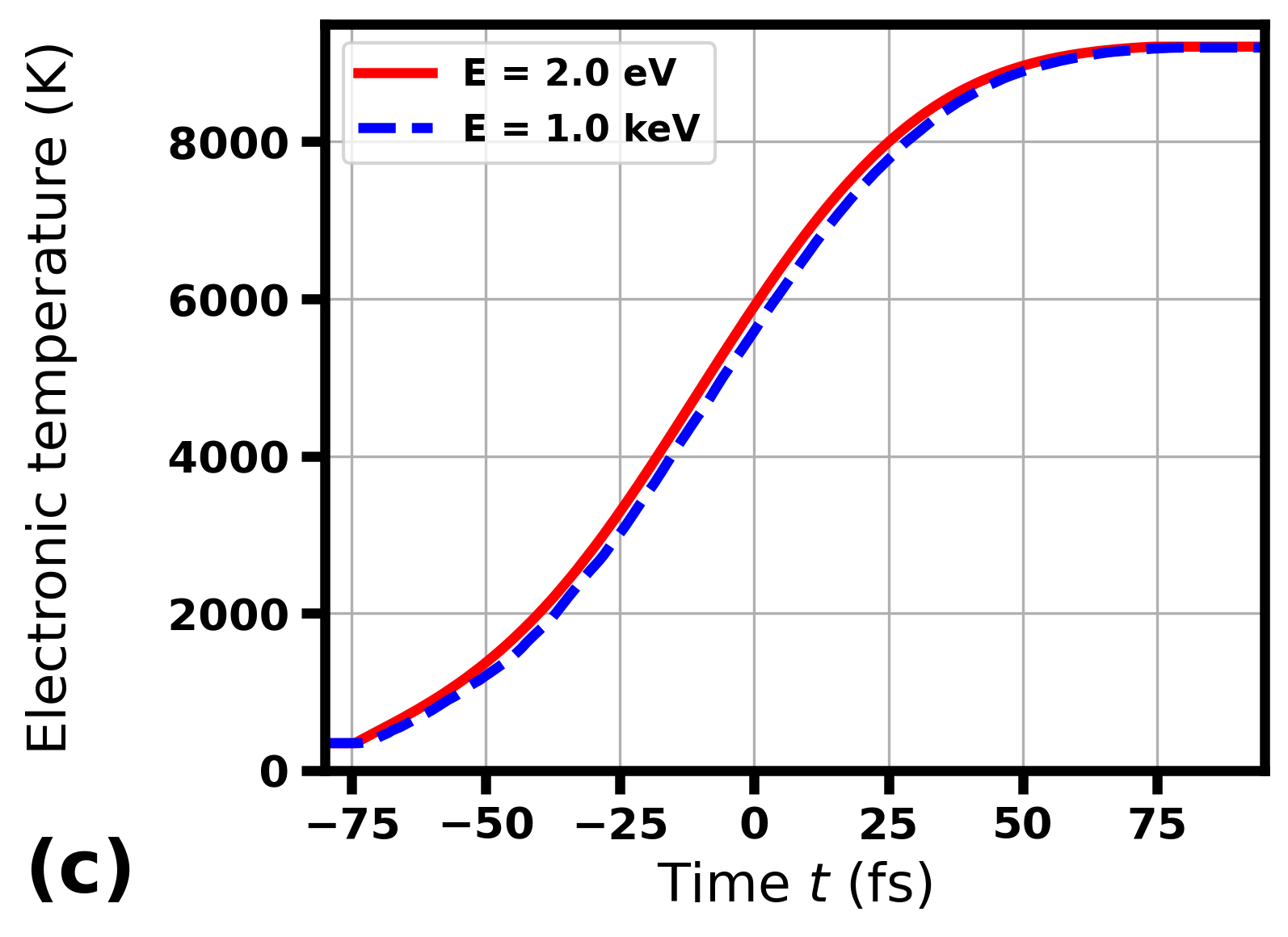}
    \includegraphics[width=\sizeonethird]{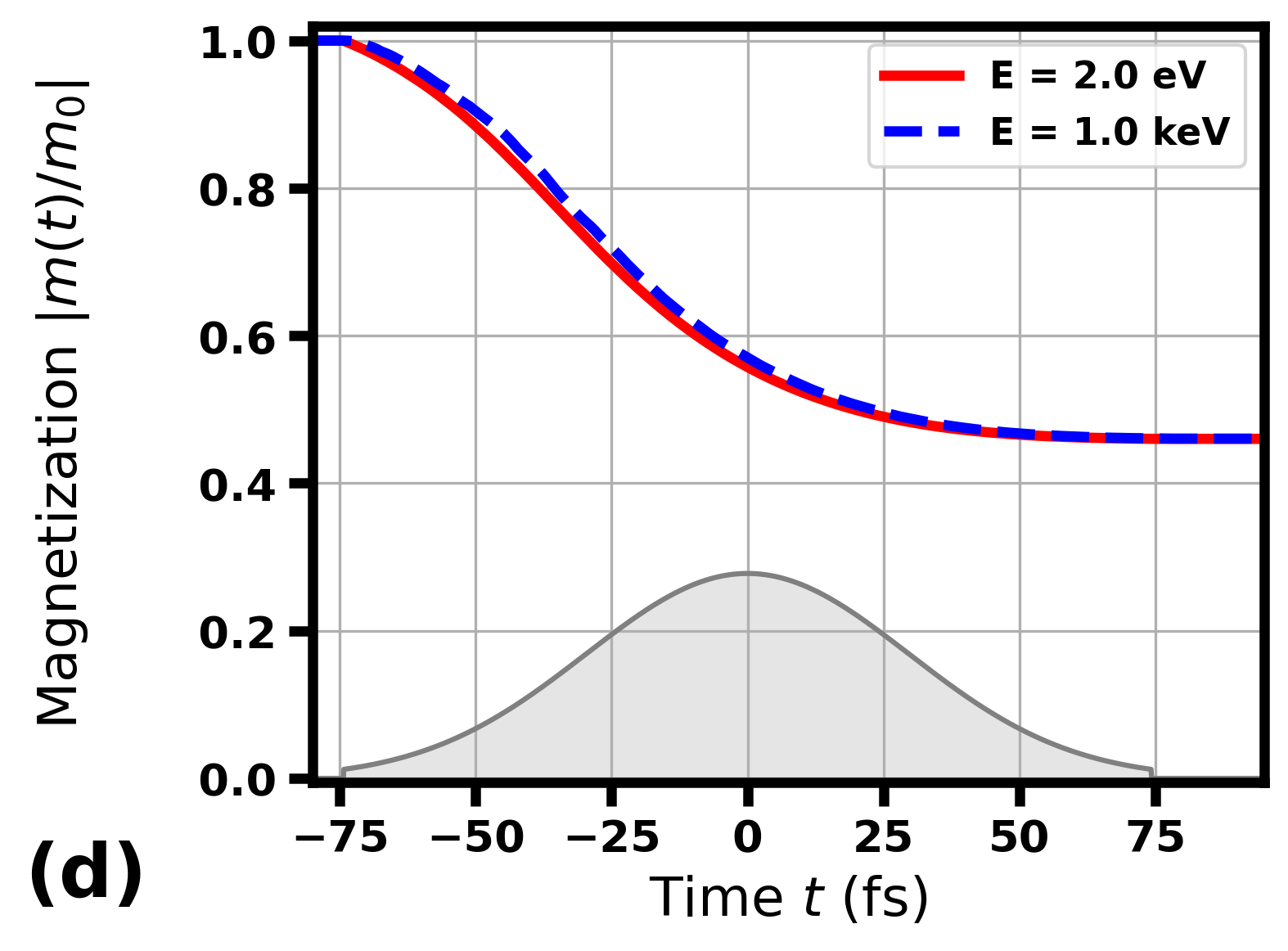}
\caption{\textbf{Transient electronic and magnetic properties of Ni predicted by XSPIN in a single magnetic domain under optical irradiation with $2$ eV pulse or X-ray irradiation with $1000$ eV pulse (as labeled), both of $70$ fs FWHM duration.}
Other simulation parameters are as in Fig.~\ref{fig1}.}
    \label{fig3}
\end{figure}
\begin{figure}[t]
    \centering
    \includegraphics[width=\sizeonethird]{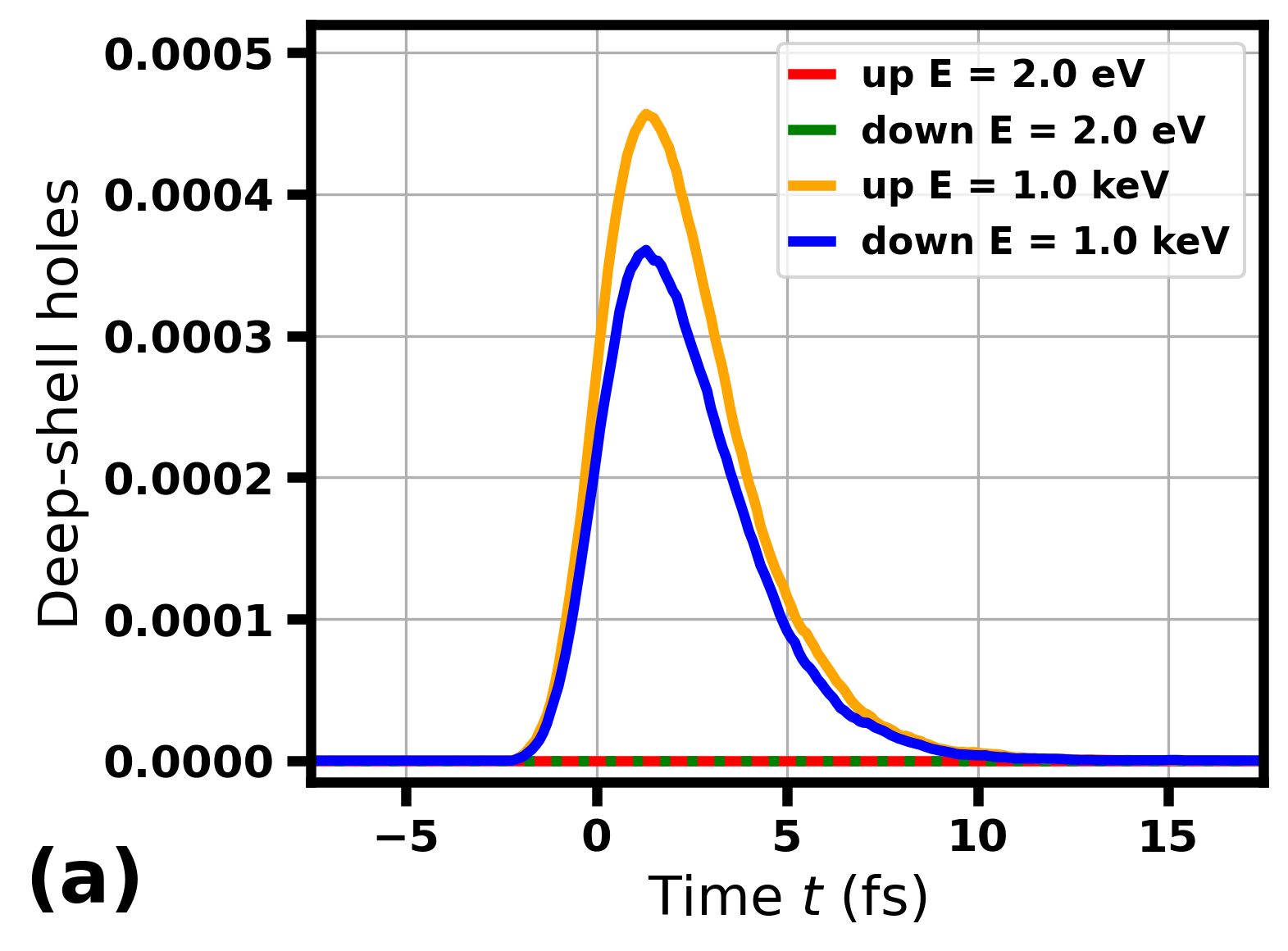}
    \includegraphics[width=\sizeonethird]{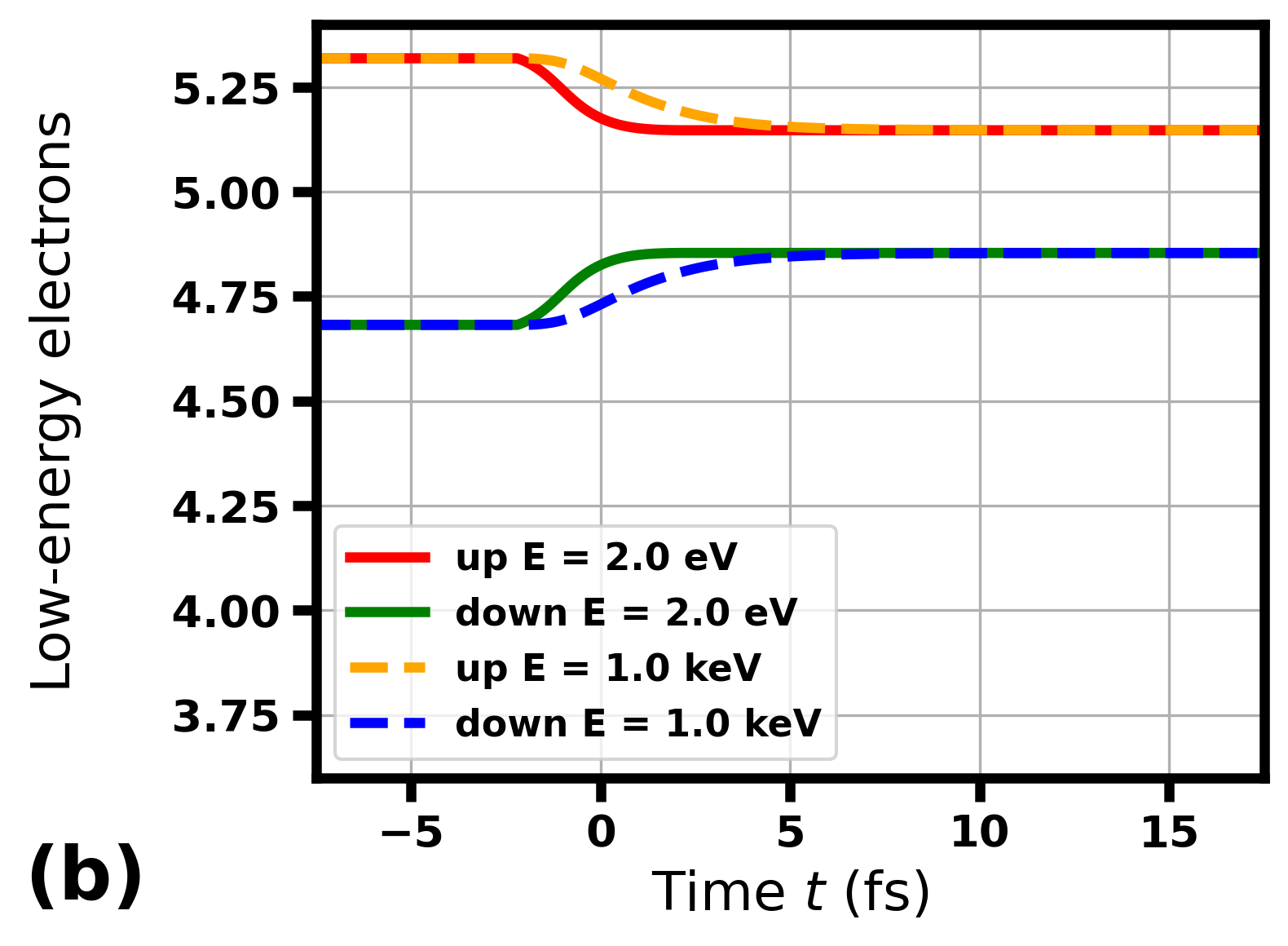}\\
    \includegraphics[width=\sizeonethird]{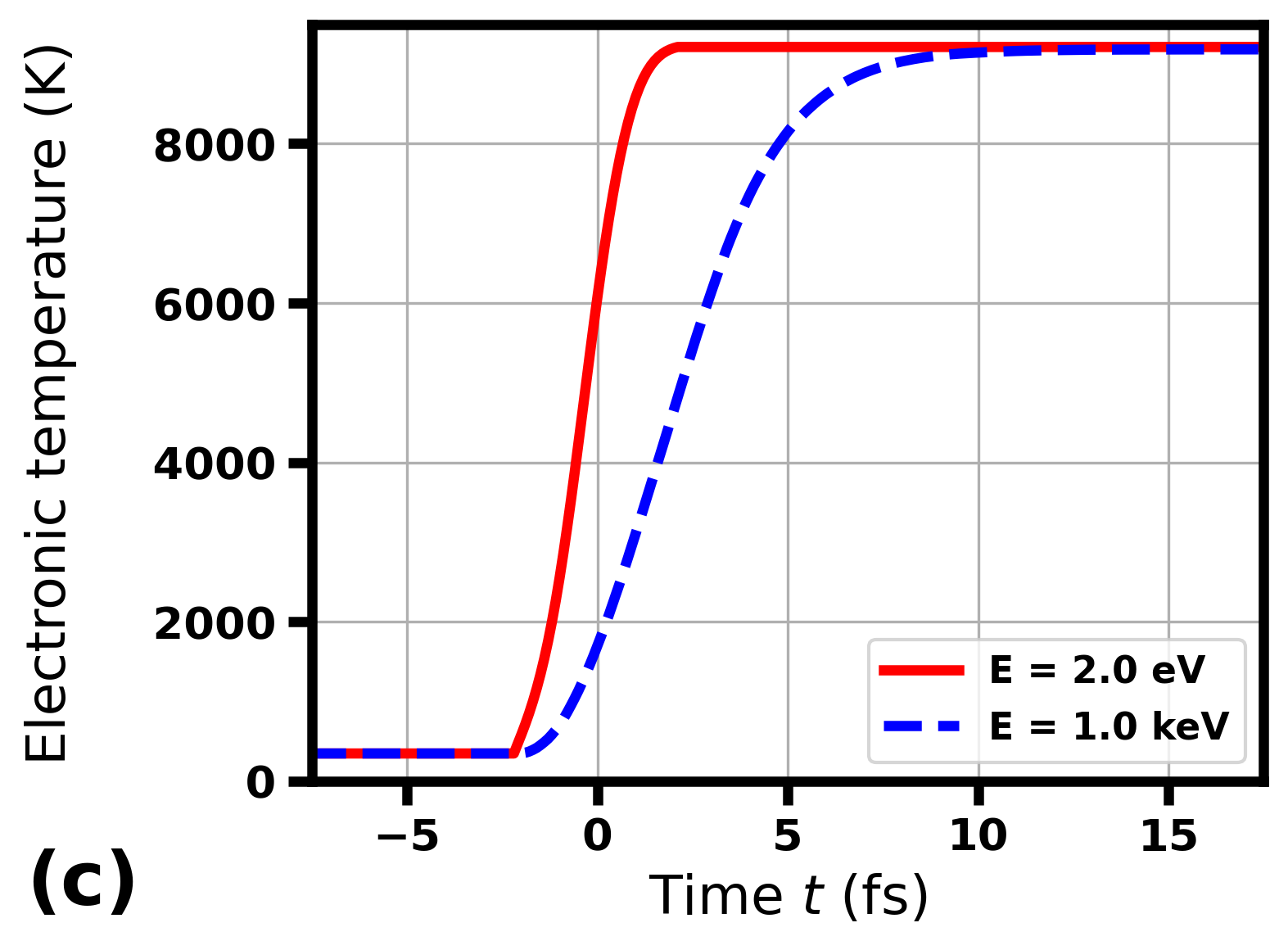}
    \includegraphics[width=\sizeonethird]{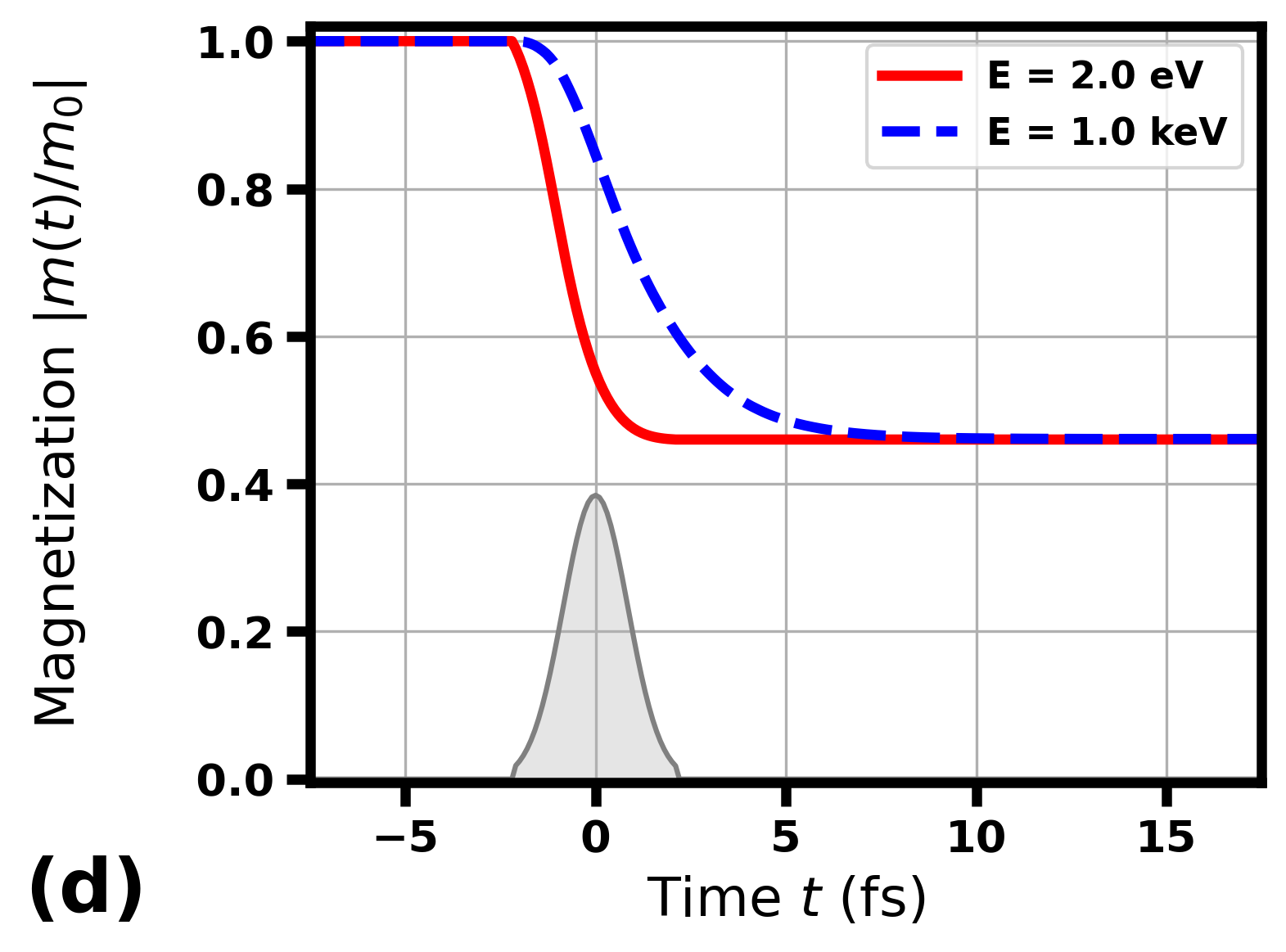}
    \caption{\textbf{Transient electronic and magnetic properties of Ni predicted by XSPIN in a single magnetic domain under optical irradiation with $2$ eV pulse or X-ray irradiation with $1000$ eV pulse (as labeled), both of $2$ fs FWHM duration.} Other simulation parameters are as in Fig.~\ref{fig1}.}
    \label{fig4}
\end{figure}

\section*{Simulation results and discussion}

We performed XSPIN simulations of the irradiated Co and Ni for a number of different energies $E=\hbar \omega_{\gamma}$ of the incoming photons, ranging from $2$ eV up to $10000$ eV, applying pulses of durations: $70$ fs FWHM or $2$ fs FWHM. The former pulse duration was used in the experiment by A.  Philippi-Kobs et al., described in \cite{KapciaNPJ2022}. The latter pulse duration, which is shorter than the relaxation time of high energy electrons, is used to investigate the impact of such a very short pulse on magnetization timescales. All simulations are performed for a fixed value of the absorbed dose corresponding to a non-destructive average absorbed dose of $0.93$ eV/atom (below the structural damage threshold; already studied in our previous papers \cite{KapciaNPJ2022,KapciaPRB2023,KapciaSciRep2024,KapciaCCC2024}). 
The dose value of $0.93$~eV/atom corresponds to different values of incoming beam fluence as the attenuation length, used for the dose-to-fluence conversion is material, photon energy and sample thickness dependent. 
In Supplementary Table S1 we provide the estimation of the fluence for the bulk systems (with sample thickness equal or larger than the attenuation length) for all cases we have studied in the manuscript (based on the data from~\cite{YuPRB1968,JohnsonPRB1974,Henke1993XRayInteractions}).
For optical and soft X-ray photons used in the work absorbed doses correspond to the beam fluence approximately between $20$ and $30$ mJ/cm$^2$.
The analyzed broad span of photon energies allows us to study impact of various electronic processes on the magnetic state of the sample. The analyzed photon energies for Co include:  $2$ eV  (with no electron impact ionization processes possible), $61.1$ eV (i.e., $1.1$ eV above the M-edge of Co; with excitation from M-edge shell possible), 
$1000$ eV and $10000$ eV  (with various deep-shell excitations possible).
For Ni, we use the same energy values, except of the energy above M-edge which is then equal to  $67.3$ eV, i.e., $1.1$ eV above the M-edge for Ni. Below we compare the results for Co and Ni samples irradiated with photons of energy $2$ eV and $1000$ eV (Figs.  \ref{fig1}--\ref{fig4}).

In Fig. \ref{fig1}, the pulse duration was $70$ fs FWHM, i.e., much longer than the cascading time of electrons produced after an impact of $1000$ eV photon ($4.0$ fs for Co), including emission of Auger electrons. The figure shows the number of deep-shell holes produced, number of low-energy electrons (i.e., those with energy below the cutoff of $15$ eV),  transient electronic temperature, and magnetization $m$ (defined as a difference between the number of spin-up and spin-down electrons in the low-energy domain, normalized to the initial value before the pulse $m_0$ at $T=300$~K),  all plotted as a function of time.  As expected, in case of $2$ eV irradiation, there are no deep-shell excitations energetically possible, whereas they are frequent in case of $1000$ eV irradiation for both spin up and spin down electrons (Fig.  \ref{fig1}a).  The production of deep-shell holes leads to a release of many energetic electrons that quickly relax, finally joining the Fermi-Dirac sea of low energy electrons (Fig.  \ref{fig1}b). As the fractions of both spin up and spin down electrons undergo a common thermalization, the increasing electronic temperature (Fig.  \ref{fig1}c) changes the populations of spin up and spin down electrons in the magnetically sensitive 3d band, and leads to an ultrafast demagnetization within the magnetic domain (Fig.  \ref{fig1}d; $m_0^{\textrm{Co}}\approx 1.65$). Notably, the above mentioned quantities (Fig.  \ref{fig1}b-d) behave in a very similar way in case of $2$ eV and $1000$ eV photon irradiation, evolving towards the same final state. 
Note that, in this work, in all calculation of the total magnetization $m$, the contributions from deep-shell holes and high-energy fraction of electrons are ignored (as their contribution to the total magnetization is negligible for all cases investigated).
 
If the cascading time of electrons released by a photon impact becomes longer than the pulse duration, one can expect some differences in the evolution of the quantities discussed above. Figs. \ref{fig2}a-d show transient number of deep-shell holes, number of low-energy electrons (i.e., those with the energy below the cutoff of $15$ eV), electronic temperature, and magnetization after irradiation of cobalt with $2$ fs FWHM pulses. The final state of all the observables is the same for the incoming photon energies of $2$ eV and $1000$ eV, however, it is delayed by several femtoseconds for $1000$ eV case, when compared to $2$ eV case. The reason for this is an extensive electron cascading process following the impact of $1000$ eV photons and the consequtive electronic relaxation, whereas in case of optical irradiation the intraband electronic relaxation is almost instantaneous. This implies that after a femtosecond optical pumping, temporal changes of electronic and magnetic parameters in a single domain in Co are mostly determined by the temporal pulse shape. Similar conclusion was drawn in \cite{KriegerJCTC2015}, however, with different demagnetization timescales resulting.

In the Supplementary Results we also show results for Co for the incoming photon energies of $\sim$ $61$ eV (above Co M-edge) and $10000$ eV (Supplementary Figs. S3 and S4, respectively). Their comparison leads to a similar conclusion as above, except of the slightly pronounced delay in demagnetization  observed for $10000$ eV case, which is due to very long cascading times for electrons released by $10000$ eV photon impact ($9.8$ fs for Co). Note also that the transient number of deep-shell holes significantly changes with the incoming photon energy due to differing photoabsorption cross-sections at different photon energies.

In order to support our findings, we have also performed computational studies of another magnetic material, Ni. It should be emphasized that cobalt and nickel differ in their basic magnetic properties, in particular, they have very different Curie temperatures,  $T_c^{\textrm{Co}}=1400$~K vs. $T_c^{\textrm{Ni}}=627$~K, respectively \cite{KapciaSciRep2024}. Figs. \ref{fig3} and \ref{fig4} show the evolution of core hole populations, electronic excitation and magnetization in a single magnetic domain of Ni after its irradiation with light pulses of $70$ fs FWHM and $2$ fs FWHM duration ($m_0^{\textrm{Ni}}\approx 0.64$). The incoming photon energies of $2$ eV and $1000$ eV are considered.  Comparison of the XSPIN predicted numbers of deep-shell holes produced, the numbers of low-energy electrons, the electronic temperatures, and magnetization curves show that also for nickel the three latter transient parameters are very close to each other, if the radiation pulse is much longer than the electron cascading time for Ni: $3.9$ fs for $1000$ eV. For a pulse duration (here $2$ fs FWHM) shorter or comparable with the electron cascading time,  demagnetization delay of several femtoseconds is again observed for $1000$ eV photon energy case and $2$ eV photon energy case before reaching the same final state. This is in agreement  with our observations for Co. We also include the results for Ni in case of photon energy close to M-edge of Ni ($67.3$ eV) and for $10000$ eV photon energy (electron cascading time $10.0$ fs ), see Supplementary Fig. S5 and S6, that are qualitatively consistent with those for Co.


\section*{Conclusions}

In summary, we performed computational studies of electronic and magnetic processes in a single magnetic domain of Co or Ni triggered by femtosecond pulses of electromagnetic radiation using our simulation tool, XSPIN \cite{KapciaNPJ2022,KapciaPRB2023,KapciaSciRep2024,KapciaCCC2024}.  These studies were mostly conducted under strongly non-equilibrium conditions. We analyzed a wide span of photon energies ranging from optical up to X-ray energies, while keeping the average absorbed dose fixed. In all analyzed cases, we predicted a sub-100 fs loss of magnetization. It was predominantly caused by the excitation of the electronic system and the subsequent redistribution of electrons within the magneto-sensitive band, as the considered timescales were too short for any phonon-mediated processes  (e.g., \cite{Nature2020}) or  processes related to inter-site Heisenberg exchange to contribute significantly (e.g., \cite{Nature2020}). The electronic excitation only depended on the absorbed radiation dose. Therefore, for a fixed absorbed dose, the final magnetization state in a single domain was the same. These predictions address those among the experimental results  \cite{BeaurepairePRL1996,KirilyukRMP2010,KoopmansNatMat2010,PfauNatCom12,SanderJPhysD17,GuttPRB10,
WangPRL2012,PfauNatCom12,MullerPRL2013a,WuPRL2016,WillemsStrDyn2017,ChenPRL2018,SchneiderPRL2020, StammNatMat07,HennesAppLSci2021} which used a few up to a few tens of femtoseconds long pulses. They qualitatively explain the similar timescales for demagnetization observed after an impact of optical or X-ray radiation (cf. Ref.~\cite{SchneiderPRL2020}). We emphasize here  that our approach does not exclude other demagnetization mechanisms that may be effective with longer radiation pulses over longer timescales. The already quite complex model is actually dedicated for application at sub-100-fs timescales, under non-equilibrium conditions. Its extension beyond this regime would require taking the slower mechanisms into account.

Generally, the measured experimental observables in a typical pump-probe experiment do not reflect truly the real timescales of the probed (de)magnetization processes, as they are intrinsically convolved with the probe beam. Depending on the probe pulse duration and the detection angle, the timescales measured in the convolved (observed) signal can be much longer than those of the real process.  Also, the temporal position of minimum in the convolved signal and the position of time zero can be significantly shifted. 
Therefore, any detailed comparison of our predictions to experimental data would require converting our results into the specific predictions for some specific observables.

In our study, we noticed that the time it takes for electrons to cascade after being released by a photon impact can influence the demagnetization timescale if the pulse duration is comparable to or smaller than the cascading time. Electron cascading can delay magnetization decrease by several femtoseconds. This mechanism can be used to control the demagnetization timescale. 

Overall, our findings contribute to a better understanding the radiation-induced magnetic response of irradiated solid materials and can be utilized in future for related practical implementations.

\section*{Acknowledgments}

The authors thank Leonard M\"uller and Andre Philippi-Kobs for helpful discussions at the early stages of the XSPIN model development.

\section*{Funding declaration}

K.J.K. thanks the Polish National Agency for Academic Exchange for funding in the frame of the National Component of the Bekker program (BPN/BKK/2022/1/00011). V.T., A.L., S.M., B.Z. acknowledge the funding received from the Collaboration Grant of the European XFEL and the Institute of Nuclear Physics, Polish Academy of Sciences. The funders had no role in the design of the study; in the collection, analyses, or interpretation of data; in the writing of the manuscript, or in the decision to publish the results.

\bibliographystyle{naturemag}

\subsection*{Data availability}

The data that support the findings of this study are available from the corresponding authors upon reasonable request.

\subsection*{Code availability}

The XSPIN code that supports the conclusions within this paper and other findings of this study is available under a license agreement. The licensor is Deutsches Elektronen-Synchrotron DESY, Notkestr. 85, 22607 Hamburg, Germany. Please contact the corresponding authors for more details.

\subsection*{Author contributions}

K.J.K and B.Z. initiated this project; K.J.K. performed all calculations with the use of  the XSPIN code, which has been developed by K.J.K and V.T.; All authors, i.e., K.J.K., V.T., F.C., A.L., S.M., P.P., and B.Z., critically discussed the results and contributed to the manuscript, which the initial version was written by K.J.K.

\subsection*{Competing interests}
Authors Victor Tkachenko, Alexander Lichtenstein,  and Sierguei Molodtsov were/are employed by the company European XFEL GmbH. Author Flavio Capotondi is employed by the company Elettra-Sincrotrone Trieste S.C.p.A. The remaining authors declare that the research was conducted in the absence of any commercial or financial relationships that could be construed as a potential conflict of interest.

\end{document}